\documentclass[onecolumn,preprint,showpacs,superscriptaddress]{revtex4-1}

\usepackage[T1]{fontenc} 
\usepackage[utf8]{inputenc}
\usepackage{mathtools}
\usepackage{amsfonts}
\usepackage{graphicx}
\usepackage{epstopdf}
\usepackage{float}
\usepackage{relsize}
\usepackage{multirow}
\usepackage{ulem}
\usepackage{comment}

\begin{document}

\title{Non-inertial effects on a non-relativistic quantum harmonic oscillator in the presence of a screw dislocation}
\author{L. C. N. Santos}
\email{luis.santos@ufsc.br}
\affiliation{Departamento de Física, CCEN-Universidade Federal da Paraíba; C.P. 5008, CEP  58.051-970, João Pessoa, PB, Brazil}
\author{F. M. da Silva}
\email{franmdasilva@gmail.com}
\affiliation{Núcleo Cosmo--ufes \& Departamento de Física, Universidade Federal do Espírito Santo, Av. Fernando Ferrari, 540, CEP 29.075-910, Vitória, ES, Brazil}
\author{C. E. Mota}
\email{clesio200915@hotmail.com}
\affiliation{Departamento de Física, CFM-Universidade Federal de Santa Catarina; C.P. 476, CEP 88.040-900, Florianópolis, SC, Brazil}
\author{V. B. Bezerra}
\email{valdirbarbosa.bezerra@gmail.com}
\affiliation{Departamento de Física, CCEN-Universidade Federal da Paraíba; C.P. 5008, CEP  58.051-970, João Pessoa, PB, Brazil}

\begin{abstract}
We investigate non-inertial effects induced by a rotating frame on a non-relativistic quantum harmonic oscillator as well as of the topology associated to a screw dislocation, which corresponds to a distortion of a vertical line into a vertical spiral. To do this, we obtain the analytical solutions of the time-independent Schrödinger equation for this harmonic oscillator potential in this background. The expressions for the energy spectrum are obtained and the solutions for four quantum states, namely $n=0,1,2$ and $3$, are analysed. Our results show that the presence of the topological defect (screw dislocation) as well the fact that we are analysing the system from the point of view of a rotating frame, changes the solutions of Schrödinger equation and the corresponding spectrum. Now these quantities depend on the angular velocity of the rotating frame, $\Omega$, and also on the parameter $\beta$, which codifies the presence of the screw dislocation. Particularly, with respect to the energy spectrum of the system the changing is such that when $\Omega$ increases, the energy can increase or decrease depending on the values we assign to the eigenvalues of the angular and linear momenta. Additionally, we observe that the values of the parameter $\beta$ that characterizes the screw dislocation causes a shift in the energy spectrum.
\end{abstract}
\pacs{03.65.Ge, 03.65.--w, 03.65.Pm, 04.20.Gz}
\maketitle

\preprint{}

\volumeyear{} \volumenumber{} \issuenumber{} \eid{identifier} \startpage{1} 
\endpage{10}

\section{Introduction}\label{sec1}

Like cracks that form when water freezes out into ice, topological defects as cosmic strings \cite{string3,string5} have an analogue interpretation which is related to phase transitions occurred in the early universe \cite{string2}. It is believed that these defects can modify the trajectory of test particles such that the energy spectrum of quantum systems can carry a dependence on the defect parameters, which characterize the spacetimes associated to them \cite{string11,santos2,string12,string13,string14,santos5,santos9}. In this context, particles under the influence of potentials that have been studied in relativistic and non-relativistic quantum mechanics in the flat spacetime with a trivial topology, can be analysed in a scenario which corresponds to spacetimes with different geometries as well as topologies, as for example, the geometries associated to disclinations and dislocations \cite{kleinert1989,katanaev1992,valanis2005material}. Indeed, this is a fruitful field of research and applications have been found in several works in the background spacetime generated to a cosmic string \cite{string12,string13,string14,santos5,santos9,lima20192d}. In which concerns similar research in the geometry associated to dislocations, several works have been performed. Among these we can mention  \cite{um,Silva2019,da2019quantum}. Particularly, in \cite{Silva2019} the authors considered the effects of the topology of the same dislocation we considering, also on an harmonic oscillator, in the framework of non-relativistic quantum mechanics, but without taking into account the angular momentum contribution to the energy spectrum arising from a system as viewed by an observer placed in a rotating frame.
In the literature, the combined effects of the topology associated with any type of topological defect and of the rotation of a frame was already considered \cite{inercial16,inercial10,santos6,dantas2015quantum,Bakkeepjplus,fonseca2016rotating,bakke2019electron}. In this work, we also consider these combined effects, taking as a system to be analysed the same considered in \cite{Silva2019}, by adding the non-inertial effects arising from a rotating frame with angular velocity $\Omega$.

Following this line of research, some well-known systems were analysed by taking into account the Klein-Gordon and Dirac equations in a generalized spacetime, providing useful insights about different systems \cite{kleinert1989}. The effects of dislocations on quantum systems have been extensively studied, by taking into account several systems. For instance, a system involving scattering of one electron by a screw dislocation with an internal magnetic flux, was considered in \cite{um}, one and two-dimensional quantum rings in the presence of a single screw dislocation \cite{dois} and quantum rings under non-inertial effects \cite{dantas2015quantum}.

On the other hand, non-inertial effects due to rotating frames is another subject of study with several interesting results. The inclusion of these effects in relativistic wave equations can be done through a coordinate transformation. The systems analysed with this scheme include the Dirac oscillator \cite{inercial10}, solutions of the DKP equation \cite{inercial15}, and the relation between the ground state energy of a scalar field and non-inertial effects \cite{inercial16,santos6}. In the non-relativistic case, the influence of a rotating frame is codified through a redefinition of Hamiltonian by assuming a coupling with the angular momentum operator, which is equivalent to introduce a rotating frame. Therefore, it is worth investigating the non-relativistic wave equations in the geometry associated with a screw dislocation, which corresponds to a distortion of a vertical line into a vertical spiral taking into account as well the non-inertial effects due to a rotating frame. In this work, we obtain the solutions of the Schrödinger equation with a harmonic oscillator potential in the background under consideration, viewed from a rotating frame and discuss in details how to express these solutions in terms of well-known functions. The energy spectrum is also obtained and some particular results are discussed.

The paper is organized as follow: In Section II we describe the line element associated to a screw dislocation and how to include non-inertial effects which is done by taking into account  an additional term in the Hamiltonian, by assuming that this procedure is equivalent to introduce a rotating frame. In Section III, we obtain the time-independent Schrödinger equation with a harmonic oscillator potential in the background under consideration. In Section IV, we find the solutions in terms of the confluent Heun function. The energy spectrum of the quantum oscillator is also obtained. Finally, in Section V the conclusions and some remarks are presented.

\section{The geometry of a screw dislocation and effect due to a rotating frame}\label{sec2}

We are committed to investigate the influence of topological defects on non-relativistic particles that experience a scalar potential $V_s$, so that the general expression for the Hamiltonian operator for this system, using the units $\hbar=1$, is given by
\begin{equation}
    \mathbb{H}_0= -\frac{1}{2m} \frac{1}{\sqrt{g^{(3)}}} \partial_i (\sqrt{g^{(3)}} g^{ij} \partial_j) + V_s,
    \label{eq01}
\end{equation}
where $g^{(3)}$ is the determinant of $g_{ij}$ and $g^{ij}$ is the contravariant metric tensor, the inverse of $g_{ij}$. Note that the indices $i$ and $j$ represent only the spatial coordinates. 

The topological defect we are going to investigate here is of the type called screw dislocation \cite{valanis2005material}, defined previously. The line element associated to the geometry of the background under consideration is written as \cite{valanis2005material}:
\begin{equation}
    ds^2 = d\rho^2 +\rho^2 d\varphi^2 +2\beta d\varphi dz +dz^2,
    \label{eq1}
\end{equation}
where $0 \leq \rho <\infty$, $-\infty < z < \infty$ and  $0\leq \varphi < 2 \pi$. The parameter $\beta$ is a positive constant that characterizes this dislocation (torsion field) \cite{Bakkeepjplus,Silva2019,daSilva2019}. At this stage we should emphasize that $\beta$ is related to the Burger vector $\vec{b}$ ($\beta\propto\left|\vec{b}\right|$) in the context of the elastic theory of solids \cite{kleinert1989,katanaev1992,valanis2005material}. According to \cite{valanis2005material}, the corresponding Burger vector is perpendicular to the plane $z=0$, and therefore, the line element (\ref{eq1}) in the context of  the elastic theory of solids describes a topological defect called a screw dislocation. Furthermore, since the value of  the parameter $\beta$ in solid state physics is of the same order as the interatomic displacement, hence, it can be defined in the range $0\,<\,\beta\,<\,1$, as discussed in Refs. \cite{Silva2019,daSilva2019}. In our results we use values for $\beta$ in the range $0\leq \beta\,<\,1$, in general, and particularly, when analysing the spectrum, we restrict to consider the following condition $\beta \ll 1$. 

Moreover, we want to analyse non-inertial effects on our system, and for that we consider that the rotating frame has a constant angular velocity $\Omega$. According to \cite{tsai1988new,dantas2015quantum,fonseca2016rotating,bakke2019electron}, the Hamiltonian operator to deal with effects of rotation on non-relativistic quantum systems is given by
\begin{equation}
    \mathbb{H} = \mathbb{H}_0 - \pmb{\Omega} \cdot \mathbb{L},
    \label{eq4.0}
\end{equation}
where $\mathbb{H}_0$ is the Hamiltonian of the system in an inertial frame and $\mathbb{L}$ is the angular momentum operator.

Thus, the transformation to a rotating frame with angular velocity $\Omega$ induces a modification in the Hamiltonian as seen from equation (\ref{eq4.0}) in the framework of non-relativistic quantum systems, introducing in this way the role played by the rotating frame through the coupling with the angular momentum operator.

\section{Quantum harmonic oscillator in a spacetime geometry with a dislocation}

We are interested to study a system subject to a harmonic oscillator potential, such that $\mathbb{H}_0$ has the following form 
\begin{equation}
    \mathbb{H}_0 \Psi= \left[ -\frac{1}{2m} \frac{1}{\sqrt{g^{(3)}}} \partial_i (\sqrt{g^{(3)}} g^{ij} \partial_j) +\frac{1}{2} m \rho^2 \omega^2 \right] \Psi,
    \label{eq4}
\end{equation}
besides that, the angular velocity is of the type $\pmb{\Omega} = \Omega \widehat{z}$, so that we also need the z-component of the angular momentum $\mathbb{L}_z$. At this point, it is interesting to call attention to the fact that, as pointed out by \cite{bezerra1997global,da2019quantum}, the presence of a dislocation will affect the component $\mathbb{L}_z$, resulting in an effective angular momentum $\mathbb{L}_z^{\text{eff}}$, given by
\begin{equation}
    \mathbb{L}_z^{\text{eff}}=-i\left( \frac{\partial}{\partial \varphi} -\beta\frac{\partial}{\partial z} \right).
    \label{eq4.1}
\end{equation}
Therefore, writing equation (\ref{eq4.0}) in the geometry given by the line element (\ref{eq1}), and considering the Hamiltonian given by equation (\ref{eq4}) plus the effective z-component of the angular momentum given by equation (\ref{eq4.1}), we obtain the following equation
\begin{equation}
    \begin{split}
    &-\frac{1}{2m} \left[ \frac{\partial^2 \Psi}{\partial \rho^2} +\frac{\rho}{\rho^2-\beta^2} \frac{\partial \Psi}{\partial \rho} +\frac{1}{\rho^2-\beta^2} \left( \frac{\partial}{\partial \varphi} -\beta \frac{\partial}{\partial z} \right)^2 \Psi +\frac{\partial^2 \Psi}{\partial z^2}  \right] \\
    &+\frac{1}{2} m \rho^2 \omega^2 \Psi -i \Omega \left( \frac{\partial}{\partial \varphi} -\beta\frac{\partial}{\partial z} \right)\Psi= E \Psi,
    \label{eq5}
    \end{split}
\end{equation}
which can be solved by the separation of variables. In this method, the wave function $\Psi$ is written in the form $\Psi\left(\rho,\,\varphi,\,z\right) = G(\rho)f(\varphi)h(z)$. Thus, let us assume that the solution is given as follows

\begin{equation}
\Psi\left(\rho,\,\varphi,\,z\right)=e^{il\varphi+ikz}\,G\left(\rho\right),
\label{eq6}
\end{equation}
where, $-\infty < k < \infty$ and $l=0, \pm 1, \pm 2, \dotsc$. Now, substituting equation (\ref{eq6}) into equation (\ref{eq5}), we obtain the following radial equation:

\begin{equation}
   G''(\rho)+\frac{\rho}{\rho^2-\beta^2}G'(\rho)-\frac{A^2}{\rho^2-\beta^2}G(\rho)-m^2 \rho^2 \omega^2 G(\rho) +B G(\rho)=0,
    \label{eq7}
\end{equation}
where we have defined the parameters $A$ and $B$ as follows
\begin{equation}
    A=l-k \beta, \qquad B=2m (E +\Omega A) -k^2.
    \label{eq8}
\end{equation}
Let us perform the change of variables $x=\rho^{2}/\beta^{2}$. Then, the radial equation (\ref{eq7}) becomes
\begin{equation}
    4x G''(x)+\frac{4x-2}{x-1}G'(x)-\frac{A^2}{x-1}G(x)-m^2 \omega^2 \beta^4 x G(x) +\beta^2 B G(x)=0.
    \label{eq9}
\end{equation}
Analysing the behaviour of equation (\ref{eq9}) when $x\rightarrow\infty$, we conclude that the solution of this equation can be written as 
\begin{equation}
G\left(x\right)=e^{-\frac{m\omega\beta^{2}}{2}\,x}\,F\left(x\right),
\label{eq10}
\end{equation}
where $F(x)$ is a function to be determined. Substituting equation (\ref{eq10}) into equation (\ref{eq9}) we obtain an equation of the form
\begin{equation}
F''(x)+\left( \alpha+ \frac{b+1}{x}+ \frac{\gamma+1}{x-1}\right)F'(x)+\left(\frac{\mu}{x}+ \frac{\nu}{x-1} \right)F(x)=0,
\label{eq11}
\end{equation}
with the parameters $\alpha$, $b$, $\gamma$, $\mu$, $\eta$, $\nu$ and $\delta$ being given by
\begin{equation}
    \alpha=-m \omega \beta^2, \qquad b=-\frac{1}{2}, \qquad \gamma=-\frac{1}{2}, \qquad \delta=\frac{1}{4}\beta^2 B, \qquad \eta=\frac{3}{8}-\frac{1}{4}A^2 -\frac{1}{4}\beta^2 B, 
    \label{eq15}
\end{equation}
\begin{equation}
    \mu= \frac{\alpha}{2}(1+b)-\frac{b}{2}(1+\gamma)-\frac{\gamma}{2}-\eta, \qquad  \nu= \frac{\alpha}{2}(1+\gamma)-\frac{b}{2}(1+\gamma)+\frac{\gamma}{2}+\eta+\delta.
    \label{eq12}
\end{equation}
As a consequence of the relations given by equations (\ref{eq15}) and (\ref{eq12}), we get the following results
\begin{equation}
    \mu= -\frac{1}{4}m \omega \beta^2+\frac{1}{4}A^2+\frac{1}{4}\beta^2 B,
    \label{eq16}
\end{equation}
and
\begin{equation}
    \nu= -\frac{1}{4}m \omega \beta^2-\frac{1}{4}A^2.
    \label{eq17}
\end{equation}

We can recognize equation (\ref{eq11}) as the confluent Heun equation \cite{ronveaux1995}, so that, $F(x)$ is given by the confluent Heun function $H_{\mathrm{C}}$
\begin{equation}
F\left(x\right)=c_1 H_{\mathrm{C}}\left(\alpha,b,\gamma,\delta,\eta,x\right),
\label{eq14}
\end{equation}
where $c_1$ is a constant to be determined by normalization procedure.

\section{Spectrum} \label{sec3}

In this section, we want to find a polynomial solution of the confluent Heun equation (\ref{eq11}). For this purpose, we start by using the Frobenius method \cite{arfken2005}, in which we write $F(x)$ as a power series around the origin as follows: $F(x) = \sum \limits_{q=0}^{\infty} a_{q}x^{q}$. Substituting this series into equation (\ref{eq11}) we obtain the following relations:
\begin{equation}
    a_0=1,
    \label{eq18}
\end{equation}
\begin{equation}
    a_1=-\frac{1}{2}(A^2 +\beta^2(B-m \omega)),
    \label{eq19}
\end{equation}
along with the three-term recurrence relation: 
\begin{equation}
    a_{q+2}=\frac{\beta^2(B-2 m \omega(1+2q))}{2(2+q)(3+2q)} a_q- \frac{\beta^2(B-m \omega(5+4q))-4(1+q)^2+A^2}{2(2+q)(3+2q)} a_{q+1}.
    \label{eq20}
\end{equation}
In order that the solution of equation (\ref{eq11}) becomes a polynomial of degree $n$, we need to truncate the series which is done by  imposing the following conditions
\begin{equation}
    a_{n+1}=0, \qquad B-2 m \omega(2n+1)=0, \qquad n=0, 1, 2, 3, \dotsc
    \label{eq21}
\end{equation}
Analysing the second condition above, we find the following result
\begin{equation}
    E_{n,k,l}=\frac{k^2}{2m}+\omega (2n+1)-\Omega A
    \label{eq22}
\end{equation}
where $E_{n,k,l}$ are the energy eigenvalues of the system under consideration. If we choose the angular frequency of the harmonic oscillator satisfying the first condition given in equation (\ref{eq21}), we conclude that this frequency depends on the quantum numbers $n$, $k$ and $l$, and thus, we will label the frequency, $\omega$, with the subscripts $n$, $k$, $l$, namely,  $\omega=\omega_{n,k,l}$. Let us start considering the case $n=0$, which requires that $a_1=0$. Thus, we find that $\omega_{0,k,l}$ is given by
\begin{equation}
    \omega_{0,k,l}=-\frac{A^2}{m \beta^2}.
    \label{eq23}
\end{equation}
We can see from the equation above that for $n=0$ the frequencies will always be negative, which is a non-physical solution and therefore will not be taken into account. Thus, the ground state for the physical system does not correspond to the state with $n=0$.

Now, let us exam the case $n=1$, in which situation we need to make $a_2=0$ and solve equation (\ref{eq21}) to $\omega_{1,k,l}$. Given that we have $\beta \ll 1$, we do not need to consider terms of $\mathcal{O}(\beta^3)$, and by doing this we find
\begin{equation}
    \omega_{1,k,l}= -\frac{A ^2 \left(A ^2-4\right)}{6 \beta ^2 m \left(A ^2-2\right)}
    \label{eq24}
\end{equation}
It is interesting to notice that the frequency of the oscillator is independent of the angular velocity of the frame, depending only on the mass, $m$, the dislocation parameter $\beta$ and on $A$. Thus, by substituting equation (\ref{eq24}) in equation (\ref{eq22}), we get that
\begin{equation}
    E_{1,k,l}=\frac{k^2}{2 m}-\frac{A ^2 \left(A ^2-4\right)}{2 \beta ^2 m \left(A ^2-2\right)}-A \Omega.
    \label{eq25}
\end{equation}
\begin{figure}[h]
    \centering
    \includegraphics[width=\textwidth]{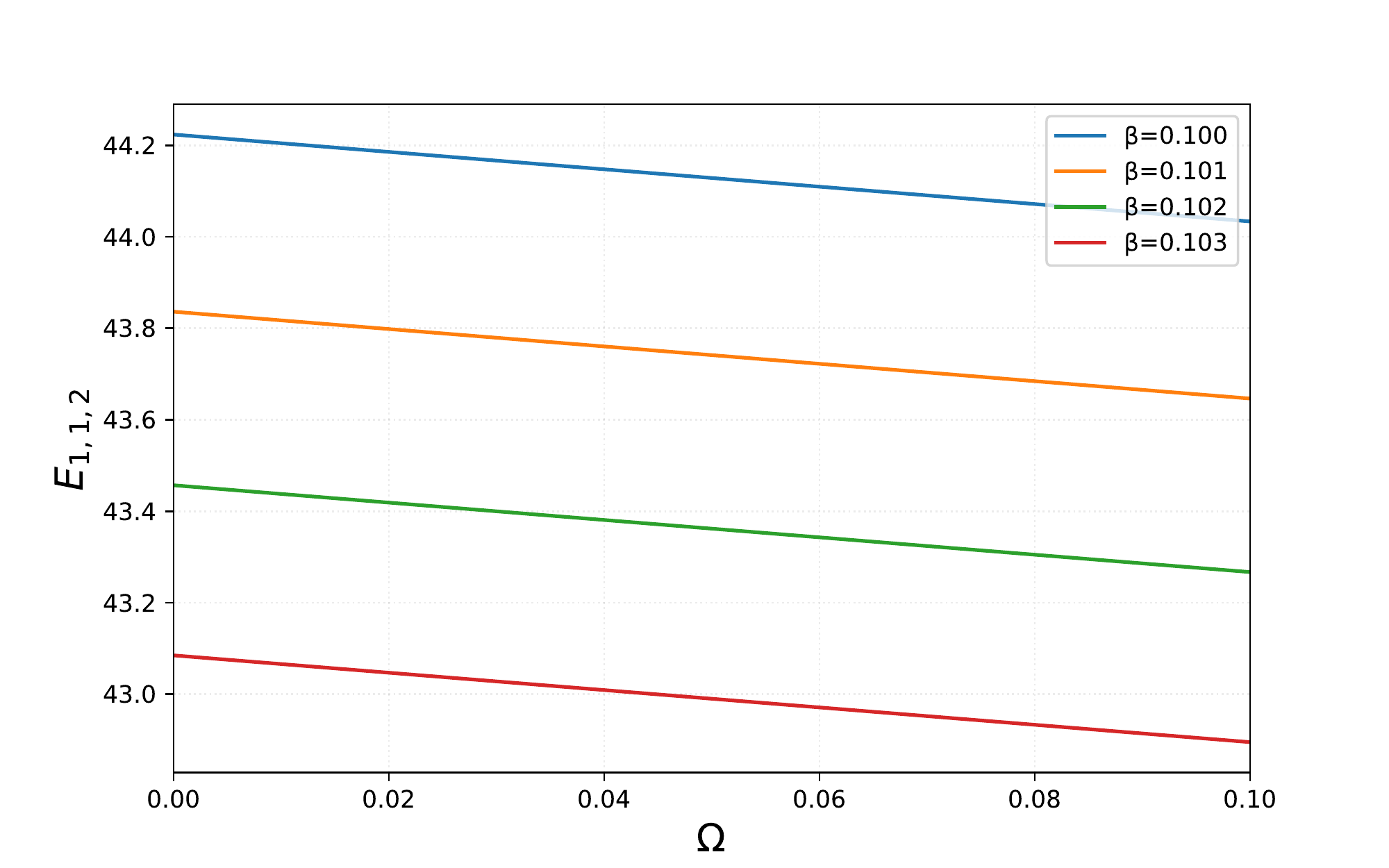}
    \caption{The energy $E_{1,1,2}$, for $n=1$, $k=1$ and $l=2$, as a function of $\Omega$ for four different values of $\beta$. Here we use $m=1$.}
    \label{fig1}
\end{figure}
In Figure \ref{fig1}, we plot the energy for the ground state $E_{1,k,l}$ as a function of the  angular velocity of the rotating frame, $\Omega$. In this figure we also see the influence of the dislocation parameter $\beta$ on the energy. We can observe that, for $l=2$ and $k=1$, when the angular velocity increases the energy decreases. However analysing equation (\ref{eq25}) we can conclude that depending on the values of $l$ and $k$ the energy for $n=1$ can increase when $\Omega$ increases. As an example, if we take $l=-2$ and $k=-1$ the energy will increase with the increasing of $\Omega$, and the frequency of the oscillator will be the same, that is $\omega_{1,1,2}=\omega_{1,-,2,-1}$. Besides, we can observe in Figure \ref{fig1} that the $E_{1,1,2}$ decreases with the increasing of the parameter $\beta$.

In Table \ref{tab1}, the frequencies  $\omega_{1,1,2}$, $\omega_{2,1,2}$ and $\omega_{3,1,2}$ of the harmonic oscillator as a function of the parameter $\beta$, are shown. We can verify that none of the frequencies depend on the rotation $\Omega$ of the frame. We also can observe that in all cases analysed, if we increase the value of $\beta$, the values of the frequencies decrease. Besides, the values of the frequencies decrease as the quantum number $n$ increases.
\begin{table}[h]
    \centering
    \begin{tabular}{ c c c c c } 
\hline
$\qquad \beta \qquad$ & $\qquad \omega_{1,1,2} \qquad$ & $\qquad \omega_{2,1,2} \qquad$ & $\qquad \omega_{3,1,2} \qquad$ \\
\hline
0.100 & 14.575 & 8.447 & 5.956 \\

0.101 & 14.445 & 8.369 & 5.900 \\

0.102 & 14.319 & 8.292 & 5.845 \\

0.103 & 14.195 & 8.218 & 5.792 \\
\hline
    \end{tabular}
    \caption{The values of $\omega_{1,1,2}$, $\omega_{2,1,2}$ and $\omega_{3,1,2}$ for four different values of $\beta$. Here $m=1$.}
    \label{tab1}
\end{table}

In what follows, let us the case $n=2$. Thus, it is necessary to make $a_3=0$ and to solve equation (\ref{eq21}) in order to obtain $\omega_{2,k,l}$. Note that, again, we will not include terms of $\mathcal{O}(\beta^3)$. By doing this, we find
\begin{equation}
    \omega_{2,k,l}=\frac{-A^6+20 A^4-64 A^2}{5 \beta ^2 m\left(3 A^4-40 A^2+64\right)}
    \label{eq26}
\end{equation}
To obtain the energy for $n=2$ we replace the equation for $\omega_{2,k,l}$ in equation (\ref{eq22}) and get
\begin{equation}
    E_{2,k,l}=\frac{k^2}{2 m}+\frac{-A^6+20 A^4-64 A^2}{\beta ^2 m\left(3 A^4-40 A^2+64\right)}-A \Omega.
    \label{eq27}
\end{equation}
\begin{figure}[h]
    \centering
    \includegraphics[width=\textwidth]{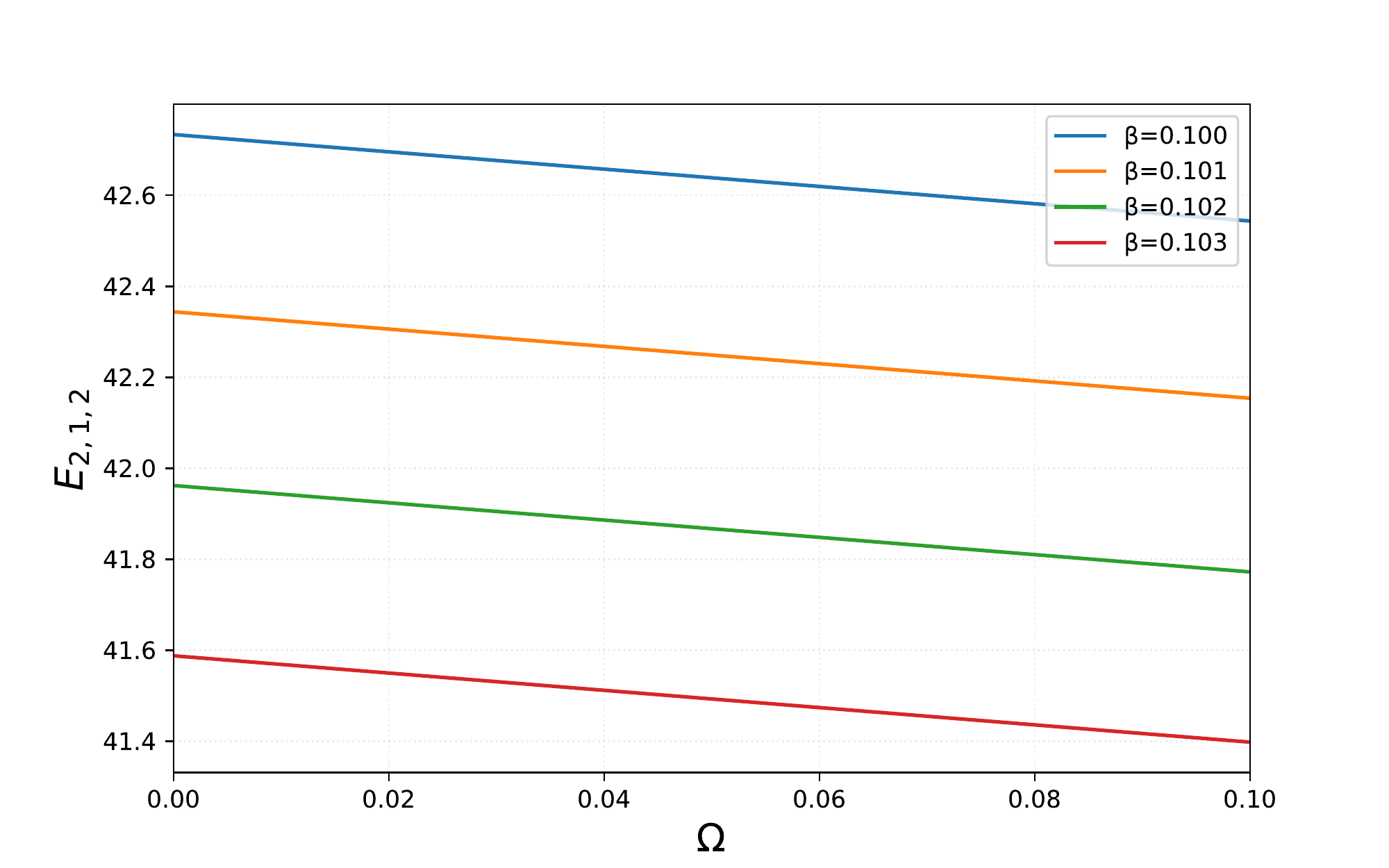}
    \caption{The energy $E_{2,1,2}$, for $n=2$, $k=1$ and $l=2$, as a function of $\Omega$ for four different values of $\beta$. Here we use $m=1$.}
    \label{fig2}
\end{figure}
In Figure \ref{fig2} we plot $E_{2,k,l}$ as a function of $\Omega$ for four different values of $\beta$. Similarly to what was observed for the ground state, when $l=2$ and $k=1$, the energy  is smaller for higher values of $\Omega$, and also smaller for higher values of the parameter $\beta$. Analysing equation (\ref{eq27}) we can see that there are values of $l$ and $k$ that make $E_{2,k,l}$ increase when $\Omega$ increases, as for example, $l=-2$ and $k=-1$. 

Proceeding in a similar way to what we did before, we can find that the values of  $\omega_{3,k,l}$ and $E_{3,k,l}$ are, respectively
\begin{equation}
    \omega_{3,k,l}=-\frac{(A-6) (A-4) (A-2) A^2 (A+2) (A+4) (A+6)}{28 \beta^2 m \left(A^6-42 A^4+392 A^2-576\right)}
    \label{eq28}
\end{equation}
and
\begin{equation}
    E_{3,k,l}=\frac{k^2}{2 m}-\frac{(A-6) (A-4) (A-2) A^2 (A+2) (A+4) (A+6)}{4 \beta^2 m \left(A^6-42 A^4+392 A^2-576\right)}-A \Omega.
    \label{eq29}
\end{equation}
\begin{figure}[h]
    \centering
    \includegraphics[width=\textwidth]{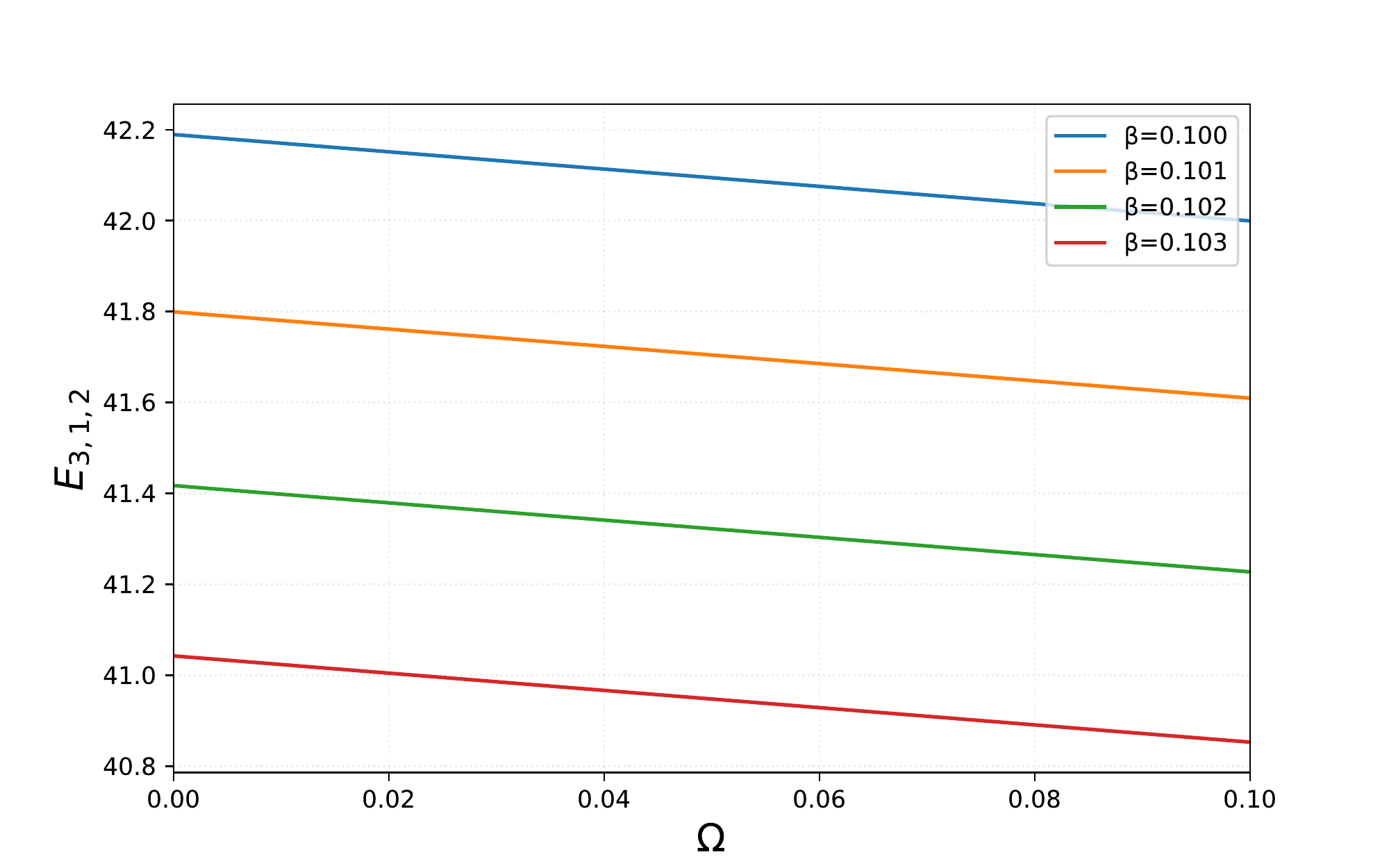}
    \caption{The energy $E_{3,1,2}$, for $n=3$, $k=1$ and $l=2$, as a function of $\Omega$ for four different values of $\beta$. Here we use $m=1$.}
    \label{fig3}
\end{figure}
In Figure \ref{fig3} we plot $E_{3,k,l}$ as a function of $\Omega$ for four different values of $\beta$. In this case, the energy also decreases with the increasing of $\Omega$ and with the increasing of $\beta$ for $l=2$ and $k=1$. Here too the behavior of $E_{3,k,l}$ with respect to the rotation of the frame can change depending on the values of $l$ and $k$. Analysing the three figures for $E_{n,k,l}$ we can conclude that the value of the energy decreases as we increase the value of the quantum number $n$, in the cases studied here.

\section{Final remarks}\label{sec4}

In this work, we have studied the behavior of non-relativistic quantum particles under influence of a harmonic oscillator potential in the background geometry generated by a screw dislocation as seen from the point of view of a non-inertial observer situated in a rotating frame. It was considered a non-relativistic wave equation in a general coordinate systems where effects of the rotation of the reference frame and effects of the topological defect considered were taken into account. We solved the resultant radial wave equation in terms of the confluent Heun function and expressed it in a polynomial form.  Due to the complexity of this type of solution, the energy spectrum of the system was investigated by considering only the first four values of the principal quantum number $n$. 

In the cases studied, we can see that the parameter associated to the angular velocity of rotation of the frame change the values of energy spectrum of the system such that when we increase the angular velocity, the energy can increase or decrease depending on the values of the eigenvalues $l$ and $k$. In addition, for the cases considered here, $E_{1,1,2}$, $E_{2,1,2}$ and $E_{3,1,2}$, we observed that the values of the parameter $\beta$ act in such a way to decrease the energy values.

It is worth calling attention to the fact that the term $A=l-k\beta$ play a hole of an effective angular momentum and carries an explicit dependence on the topology associated to the defect. Some systems with a magnetic flux through a solenoid present the so-called Aharonov--Bohm effect for bound states. This type of effect can be associated to a term $l_{eff}=l-e\Phi/2\pi$ (with $\Phi$ being the magnetic flux and $e$ the electric charge \cite{bezerra1997global}). As observed in the literature \cite{Bakkeepjplus,bakke2019electron,bezerra1997global}, by comparing this relation with the one obtained in this work, we can see that the term associated to the screw dislocation $\beta$ generates an analogous effect.


\section{Acknowledgments}
CEM has a scholarship paid by Capes (Brazil), LCNS would like to thank Conselho Nacional de Desenvolvimento Cient\'ifico e Tecnol\'ogico (CNPq) for partial financial support through the research Project No. 164762/2020-5 and FMS would like to thank CNPq for financial support through the research Project No. 165604/2020-4. VBB is partially supported by CNPq through the Research Project No. 307211/2020-7.

\bibliographystyle{ieeetr}
\bibliography{referencias_unificadas}

\begin{thebibliography}{10}

\bibitem{string3}
A.~Vilenkin, ``Gravitational field of vacuum domain walls and strings,'' {\em
  Phys. Rev. D}, vol.~23, pp.~852--857, Feb 1981.

\bibitem{string5}
W.~A. Hiscock, ``Exact gravitational field of a string,'' {\em Phys. Rev. D},
  vol.~31, pp.~3288--3290, Jun 1985.

\bibitem{string2}
A.~Vilenkin, ``Cosmic strings,'' {\em Phys. Rev. D}, vol.~24, pp.~2082--2089,
  Oct 1981.

\bibitem{string11}
S.~{\"O}lmez, V.~Mandic, and X.~Siemens, ``Gravitational-wave stochastic
  background from kinks and cusps on cosmic strings,'' {\em Phys. Rev. D},
  vol.~81, no.~10, p.~104028, 2010.

\bibitem{santos2}
L.~C.~N. Santos and C.~C. Barros~Jr., ``Scalar bosons under the influence of
  noninertial effects in the cosmic string spacetime,'' {\em Eur. Phys. J. C},
  vol.~77, no.~3, p.~186, 2017.

\bibitem{string12}
A.~L. Cavalcanti~de Oliveira and E.~R. Bezerra~de Mello, ``{Exact solutions of
  the Klein-Gordon equation in the presence of a dyon, magnetic flux and scalar
  potential in the specetime of gravitational defects},'' {\em Class. Quant.
  Grav.}, vol.~23, pp.~5249--5264, 2006.

\bibitem{string13}
E.~R. Figueiredo~Medeiros and E.~R.~B. de~Mello, ``{Relativistic quantum
  dynamics of a charged particle in cosmic string spacetime in the presence of
  magnetic field and scalar potential},'' {\em Eur. Phys. J.}, vol.~C72,
  p.~2051, 2012.

\bibitem{string14}
M.~Hosseinpour, F.~M. Andrade, E.~O. Silva, and H.~Hassanabadi, ``Scattering
  and bound states for the hulth{\'e}n potential in a cosmic string
  background,'' {\em Eur. Phys. J. C}, vol.~77, p.~270, Apr 2017.

\bibitem{santos5}
L.~C.~N. Santos and C.~C. Barros~Jr., ``Relativistic quantum motion of spin-0
  particles under the influence of noninertial effects in the cosmic string
  spacetime,'' {\em Eur. Phys. J. C}, vol.~78, p.~13, Jan 2018.

\bibitem{santos9}
F.~A.~C. Neto, F.~M. Da~Silva, L.~C.~N. Santos, and L.~B. Castro, ``{Scalar
  bosons with Coulomb potentials in a cosmic string background: Scattering and
  bound states},'' {\em Eur. Phys. J. Plus}, vol.~135, no.~1, p.~25, 2020.

\bibitem{kleinert1989}
H.~Kleinert, {\em Gauge Fields in Condensed Matter, Vol. II--Stresses and
  Defects}.
\newblock World Scientific, 1989.

\bibitem{katanaev1992}
M.~O. Katanaev and I.~V. Volovich, ``{Theory of defects in solids and
  three-dimensional gravity},'' {\em Annals Phys.}, vol.~216, pp.~1--28, 1992.

\bibitem{valanis2005material}
K.~Valanis and V.~Panoskaltsis, ``Material metric, connectivity and
  dislocations in continua,'' {\em Acta mechanica}, vol.~175, no.~1,
  pp.~77--103, 2005.

\bibitem{lima20192d}
D.~F. Lima, F.~M. Andrade, L.~B. Castro, C.~Filgueiras, and E.~O. Silva, ``On
  the 2d dirac oscillator in the presence of vector and scalar potentials in
  the cosmic string spacetime in the context of spin and pseudospin
  symmetries,'' {\em The European Physical Journal C}, vol.~79, no.~7,
  pp.~1--15, 2019.

\bibitem{um}
C.~Furtado, V.~B. Bezerra, and F.~Moraes, ``{Quantum scattering by a magnetic
  flux screw dislocation},'' {\em Phys. Lett. A}, vol.~289, pp.~160--166, 2001.

\bibitem{Silva2019}
W.~C.~F. da~Silva, K.~Bakke, and R.~L.~L. Vit\'oria, ``{Non-relativistic
  quantum effects on the harmonic oscillator in a spacetime with a distortion
  of a vertical line into a vertical spiral},'' {\em Eur. Phys. J. C}, vol.~79,
  no.~8, p.~657, 2019.

\bibitem{da2019quantum}
W.~da~Silva and K.~Bakke, ``Quantum aspects of a quantum particle in a
  cylindrical wire in the presence of a screw dislocation,'' {\em The European
  Physical Journal Plus}, vol.~134, no.~4, pp.~1--6, 2019.

\bibitem{inercial16}
H.~F. Mota and K.~Bakke, ``{Noninertial effects on the ground state energy of a
  massive scalar field in the cosmic string spacetime},'' {\em Phys. Rev. D},
  vol.~89, no.~2, p.~027702, 2014.

\bibitem{inercial10}
K.~Bakke, ``{Rotating effects on the Dirac oscillator in the cosmic string
  spacetime},'' {\em Gen.Rel.Grav.}, vol.~45, pp.~1847--1859, 2013.

\bibitem{santos6}
L.~C.~N. Santos and C.~C. Barros~Jr., ``Rotational effects on the casimir
  energy in the space-time with one extra compactified dimension,'' {\em Int.
  J. Mod. Phys. A}, vol.~33, p.~1850122, 07 2018.

\bibitem{dantas2015quantum}
L.~Dantas, C.~Furtado, and A.~S. Netto, ``Quantum ring in a rotating frame in
  the presence of a topological defect,'' {\em Physics Letters A}, vol.~379,
  no.~1-2, pp.~11--15, 2015.

\bibitem{Bakkeepjplus}
K.~Bakke, ``{Effects of rotation in the spacetime with the distortion of a
  vertical line into a vertical spiral},'' {\em Eur. Phys. J. Plus}, vol.~134,
  no.~11, p.~546, 2019.

\bibitem{fonseca2016rotating}
I.~Fonseca and K.~Bakke, ``Rotating effects on the landau quantization for an
  atom with a magnetic quadrupole moment,'' {\em The Journal of chemical
  physics}, vol.~144, no.~1, p.~014308, 2016.

\bibitem{bakke2019electron}
K.~Bakke, R.~Ribeiro, and C.~Salvador, ``On an electron in a nonuniform axial
  magnetic field in a uniformly rotating frame,'' {\em International Journal of
  Modern Physics A}, vol.~34, no.~33, p.~1950229, 2019.

\bibitem{dois}
A.~S. Netto, C.~Chesman, and C.~Furtado, ``Influence of topology in a quantum
  ring,'' {\em "Phys. Lett. A"}, vol.~372, no.~21, pp.~3894--3897, 2008.

\bibitem{inercial15}
M.~Hosseinpour and H.~Hassanabadi, ``{DKP Equation in a Rotating Frame with
  Magnetic Cosmic String Background},'' {\em Eur. Phys. J. Plus}, vol.~130,
  no.~11, p.~236, 2015.

\bibitem{daSilva2019}
W.~C.~F. da~Silva and K.~Bakke, ``{Non-relativistic effects on the interaction
  of a point charge with a uniform magnetic field in the distortion of a
  vertical line into a vertical spiral spacetime},'' {\em Class. Quant. Grav.},
  vol.~36, no.~23, p.~235002, 2019.

\bibitem{tsai1988new}
C.-H. Tsai and D.~Neilson, ``New quantum interference effect in rotating
  systems,'' {\em Physical Review A}, vol.~37, no.~2, p.~619, 1988.

\bibitem{bezerra1997global}
V.~B. Bezerra, ``Global effects due to a chiral cone,'' {\em Journal of
  Mathematical Physics}, vol.~38, no.~5, pp.~2553--2564, 1997.

\bibitem{ronveaux1995}
A.~Ronveaux, {\em {Heun’s Differential Equations}}.
\newblock Oxford: Oxford University Press, 1995.

\bibitem{arfken2005}
G.~Arfken, {\em {Mathematical Methods for Physicists}}.
\newblock New York: 6th edn. Elsevier Academic Press, 2005.

\end{thebibliography}

$\bigskip $

\end{document}